# Reorientation kinetics of superparamagnetic nanostructured rods

J. Fresnais and J.-F. Berret[@]

Matière et Systèmes Complexes, UMR 7057 CNRS Université Denis Diderot Paris-VII, Bâtiment Condorcet, 10 rue Alice Domon et Léonie Duquet, 75205 Paris, France

B. Frka-Petesic, O. Sandre and R. Perzynski

UPMC Univ Paris 06 - Laboratoire Liquides Ioniques et Interfaces Chargées, UMR 7612 CNRS, 4 place Jussieu-case 63, F-75252 Paris Cedex 05 France

**Abstract**
The attractive interactions between oppositely charged species (colloids, macromolecules etc) dispersed in water are strong, and the direct mixing of solutions containing such species generally yields to a precipitation, or to a phase separation. We have recently developed means to control the electrostatically-driven attractions between nanoparticles and polymers in water, and at the same time to preserve the stability of the dispersions. We give here an account of the formation of supracolloidal aggregates obtained by co-assembly of 7 nm particles with copolymers. Nanostructured rods of length comprised between 5 and 50 μm and diameter 500 nm were investigated. By application of a magnetic field, the rods were found to reorient along with the magnetic field lines. The kinetics of reorientation was investigated using step changes of the magnetic field of amplitude $\pi/2$. From the various results obtained, among which an exponential decay of the tangent of the angle made between the rod and the field, we concluded that the rods are superparamagnetic.

[@] : jean-francois.berret@univ-paris-diderot.fr

# I - INTRODUCTION

Since the pioneering work by Kataoka and Harada [1, 2], it has been recognized that attractive interactions between polyelectrolyte-neutral diblock copolymers and oppositely charges species resulted in the formation of stable colloids. These colloids were found to form spontaneously by electrostatic co-assembly, and their microstructure was identified as being of core-shell type. The cores were described as dense coacervate microphases comprising the oppositely charged species, whereas the corona was made from the neutral blocks and surrounding the cores. With polyelectrolyte-neutral copolymers, the complexation is controlled by the appropriate choices of the polymer, of its molecular weight and by the molecular weight ratio between the two blocks. The specimens examined so far comprise synthetic [3, 4] and biological [1, 5] macromolecules, multivalent counterions [6], surfactant micelles [7-9]. The formation of the mixed aggregates is generally understood as the result of a nucleation and growth mechanism of a microphase made from the oppositely charged constituents. This growth is arrested at a size which is fixed by the dimension of the polymer, *i.e.* the range 20 – 200 nm [8]. Recently, other scenarios of controlled clustering of particles have been elaborated, and in particular scenarios dealing with the design of "colloidal molecules" in the 100 nm – 1 µm range [10-15].

Here, we report on the electrostatic complexation between anionically charged iron oxide nanoparticles and oppositely charged block copolymers, and demonstrate that it is effective for the elaboration of novel types of nanostructured colloids [16-18]. We have investigated anionically modified dispersions of iron oxide particles that were stabilized by poly(acrylic acid) [19, 20]. By combining an appropriate choice of copolymer [17] and of kinetics of association [21], we were able to co-assemble 7 nm particles into spherical or cylindrical aggregates in the micrometer range. For these assemblies, it was also shown that the particles were aggregated in a densely packed state, with an internal volume fraction of 30 %. In the present communication, only the case of nanostructured rods will be discussed. Here, we derived the probability distribution function of length and studied their mechanisms of reorientations under the application of a magnetic field. The experimental observations lead us to the conclusion that rods are superparamagnetic.

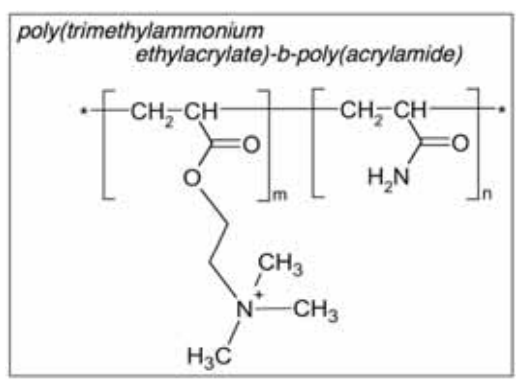

***Figure 1*** : *Chemical structure of the diblock copolymer PTEA$_{11K}$-b-PAM$_{30K}$ investigated in the present work. The abbreviation PTEA stands for poly(trimethylammonium ethylacrylate methylsulfate) and PAM for poly(acrylamide).*

# II - Experimental

## II.1 - Polymers

The anionically charged nanoparticles have been complexed with a cationic-neutral diblock copolymer, referred to as poly(trimethylammonium ethylacrylate)-b-poly(acrylamide) (Fig. 1). The counterion associated with the quaternary ammonium group was methyl sulfate. The diblock copolymers were synthesized by MADIX® controlled radical polymerization which is a Rhodia patented process [22, 23]. The molecular weight put under scrutiny in this study corresponds to m = 41 monomers in the charged blocks ($M_W$ = 11 000 g mol$^{-1}$) and n = 420 for the neutral chain ($M_W$ = 30 000 g mol$^{-1}$). In accordance with previous reports [8, 9], the copolymer was abbreviated PTEA$_{11K}$-$b$-PAM$_{30K}$. In aqueous solutions at neutral pH, the chains are dispersed and in the state of unimers. Light scattering performed in the dilute regime have revealed a molecular weight $M_W$ = 35 000 ± 2000 g mol$^{-1}$ and an hydrodynamic diameter $D_H$ = 11 ± 1 nm [8]. The polydispersity index was determined by size exclusion chromatography at 1.6.

## II.2 - Iron Oxide Nanoparticles

The synthesis of magnetic nanoparticles investigated here was elaborated by R. Massart three decades ago using the technique of « soft chemistry » [24]. This technique is based on the polycondensation of metallic salts in alkaline aqueous media, resulting in the formation of magnetite ($Fe_3O_4$) nanoparticles of sizes comprised between 4 and 15 nm. Magnetite is further oxidized into maghemite ($\gamma$-$Fe_2O_3$) and the nanoparticles are sorted according to their size. Fig. 2 displays typical transmission electron cryo-microscopy (cryo-TEM) images of $\gamma$-$Fe_2O_3$ nanoparticles [16]. These dispersions have been studied thoroughly with respect to several fundamental properties, including the size distribution [25], the modification of the surface chemistry of the particles [26] and the magnetic birefringence and superparamagnetism, which described the behavior of the particles in magnetic field and gradients [27].

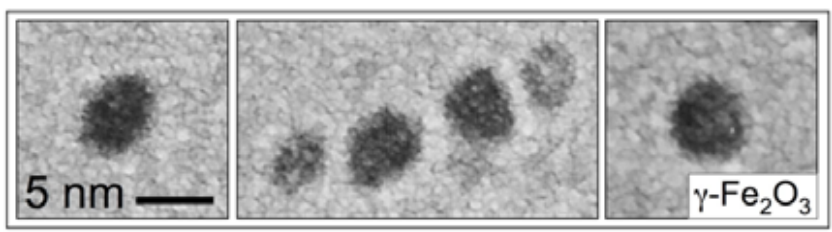

*Figure 2 : Iron oxide nanoparticles as observed by transmission electron cryo-microscopy [16]. The stability of the dispersion was ensured by electrostatic interactions mediated by the native cationic charges, or by oligomers adsorbed on their surfaces.*

In the conditions of the synthesis (pH 1.8, weight concentration c = 6.58 wt. %, volume fraction of $\phi$ = 1.29 %), the magnetic dispersions were thermodynamically stable over a period of years. They were stabilized by electrostatic interactions arising from the native cationic charges at the surface of the particles. Nitrate ions served as counterions and insured electroneutrality.

Vibrating sample magnetometry consisted in measuring the magnetization *versus* excitation M(H) for a solution at volume fraction $\phi$. Fig. 3 shows the evolution of the macroscopic magnetization M(H) normalized by its saturation value $M_S$ for the present $\gamma$-$Fe_2O_3$ batch. Here, $M_S = \phi m_S$, where $m_S$ is the volumetric magnetization of maghemite ($m_S = 2.6 \times 10^5$ A m$^{-1}$). The solid curve in Fig. 3 was obtained using the Langevin function for paramagnetism convoluted with a log-normal distribution function of the particle size. The parameters of the distribution are the median diameter ($\tilde{D}_{NP}$ = 7.1 ± 0.3 nm) and the polydispersity (s = 0.26 ± 0.03). With dynamic light scattering, a single mode in the autocorrelation function was observed, corresponding to a hydrodynamic diameter $D_H$ = 14 ± 1 nm and a polydispersity index of 0.12. The cationic particles were further coated by poly(acrylic acid) oligomers with molecular weight 2000 g mol$^{-1}$, using the precipitation-redispersion process described previously [19]. The hydrodynamic sizes found in $\gamma$-$Fe_2O_3$-$PAA_{2K}$ dispersions were 4 nm above that of the bare particles. This 4 nm-increase was assigned to the presence of a strongly charged $PAA_{2K}$ brush around the particles.

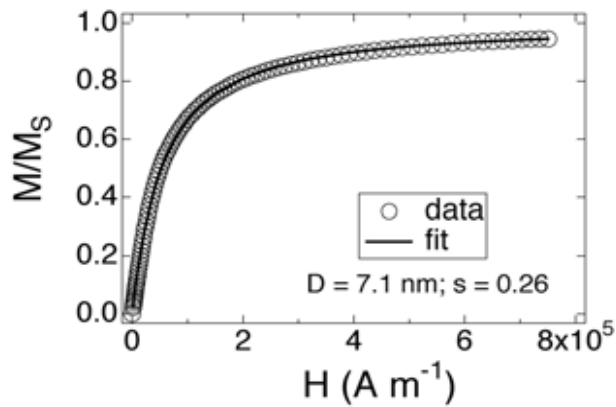

*Figure 3* : *Magnetic field dependence of the macroscopic magnetization M(H) normalized by its saturation value $M_S$ for cationic maghemite dispersions. The solid curve was obtained using the Langevin function for paramagnetism convoluted with a log-normal distribution function for the particle sizes, given with a median diameter $\tilde{D}_{NP}$ = 7.1 nm and a polydispersity 0.26.*

### II.3 - Optical microscopy
Phase-contrast images of the rods were acquired on a IX71 inverted microscope (Olympus) equipped with a 100X objective. We used a Photometrics Cascade camera (Roper Scientific) and Metaview software (Universal Imaging Inc.). For the observations of the reorientations under external magnetic field, we used a Leitz (Ortholux) upward microscope with a ×20 objective. The magnetic field was applied using two pairs of coils orthogonal to each other in a set-up described earlier [28]. Video sequences were recorded by a CCD camera, digitized and treated by the ImageJ software (http://rsbweb.nih.gov/ij/).

*II.4 – Fabrication of the rods*
The protocols for the elaboration of nanostructured rods have been reported recently [21]. Here, we provide a brief account of the exploited procedures. Solutions of $PTEA_{11K}$-$b$-$PAM_{30K}$ and of $PAA_{2K}$-coated iron oxide nanoparticles were first prepared in 1 M ammonium chloride ($NH_4Cl$) at weight concentration c = 0.2 wt. %. The two solutions were then mixed, yielding a disperse solution where polymers and particles were not yet associated. The absence of interactions between particles and polymers at high salt content was carefully checked by dynamic light scattering. The ratio of the electrostatic charges borne by the particles and by the polymers was fixed at the charge stoichiometry. The electrostatic interactions were then monitored by a slow removal of the salt by dialysis. Dialysis was performed using a slide-a-lyzer cassette with 10 kD molecular weight cut-off. For the elaboration of the rods, dialysis was operated under a constant magnetic field of 0.1 T. Once the ionic strength of the dialysis bath has reached its stationary value, typically
$10^{-2}$ M, the magnetic field was removed and the solutions were studied by different techniques, including optical microscopy.

## III - Results

*Optical microscopy* : Fig. 4a shows an image (×100) of a rod dispersion sealed between glass plates. In the conditions of the figure, no magnetic field was applied. Elongated structures with random orientations were clearly visible, with typical sizes in the micrometer range. For this specimen, an image analysis was carried out so as to derive the length distribution of the rods. Fig. 5 displays the probability distribution function observed by microscopy on a series of images similar to that of Fig. 4a. The data could be fitted using a log-normal function with median length $\tilde{L}_{Rod}$ = 12.3 ± 0.7 μm, and a polydispersity s = 0.55. Fig. 4a deserves two additional comments : *i)* the rods are characterized by a persistence length that is large, presumably larger than 100 μm [21] and *ii)* they did not display signs of destabilization, even after prolonged period under high magnetic field. As already mentioned, in absence of magnetic field, the rods in Fig. 4a were found to be randomly oriented. However, if a magnet was brought near to the glass plate, the rods reoriented spontaneously and followed the magnetic field lines. Fig. 4b illustrates orientations of the rods in the plane of the figure. Note finally in Fig. 4b that even under the application of a field, the rods remained dispersed and unaggregated. The origin of the coupling between the rod and the external field is examined below.

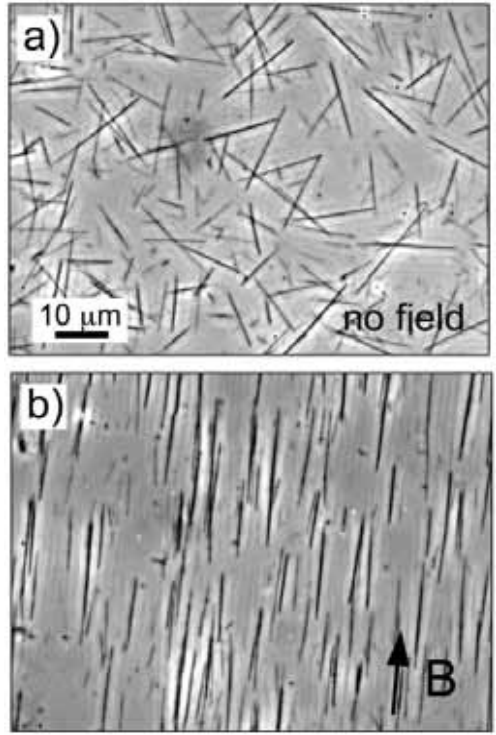

**Figure 4** : *Phase contrast optical microscopy images (×100) of a dispersion of nanostructured rods without magnetic field (a), and with a magnetic field applied in the plane of observation (b).*

**π/2-reorientation kinetics** : We have performed quantitative measurements of the kinetics associated with π/2-reorientations using the microscopy set-up described in the experimental Section. As already mentioned, the magnetic fields were applied using two pairs of coils orthogonal to each other. An isolated rod located far from the cell walls was selected and oriented by the first pair of coils. Once aligned, this primary field was switched off, and the perpendicular field was switched on thanks to the second set of coils. Experiments were carried out for rod lengths 5 – 50 μm and magnetic excitations $10^3$ - $10^4$ A m$^{-1}$.

In the following, θ(t) denotes the angle between the major axis of the selected rod and the applied excitation H. An illustration of a π/2-reorientation is shown in the inset of Fig. 6. With the conventions defined in the inset, the rod was initially parallel to the X-axis. In its final position, it was aligned with respect to the Y-axis. The rod reorientation kinetics was investigated systematically, yielding three main observations :

*i)* At the application of the field, the rod did not respond instantaneously. On the contrary, the rod seemed to "hesitate" around the initial orientation $\theta_I = \pi/2$. After a delay time noted $t_D$, the rod then rotated in the plane in a propeller-like motion up to its final position. Fig. 6 displays the time dependence of the angle $\theta(t)$ for a rod of length $L_{Rod} = 8.5$ μm. There, the applied field was $5.6 \times 10^3$ A m$^{-1}$. The delay time $t_D$ (= 0.55 s) for this experiment is indicated by an arrow.

*ii)* Starting from their initial orientation, the rods could go either clockwise or counterclockwise with an equi-probability, yielding for the final angles either $\theta_F = 0$ or to $\theta_F = \pi$.

*iii)* The relaxation of the rod was analyzed in terms of tg$\theta$ *versus* time. Figs. 7a and 7b illustrate the final relaxations for two different rods with length $L_{Rod} = 10.7$ μm and $L_{Rod} = 23.6$ μm, respectively. The time was counted from the delay time $t_D$ defined in *i)*. Under these conditions, tg$\theta(t)$ was found to decrease exponentially with time, as indicated by the straight lines in the semi-logarithmic representation. The decrease of the reorientation was found to strongly depend on the strength of the magnetic excitation H.

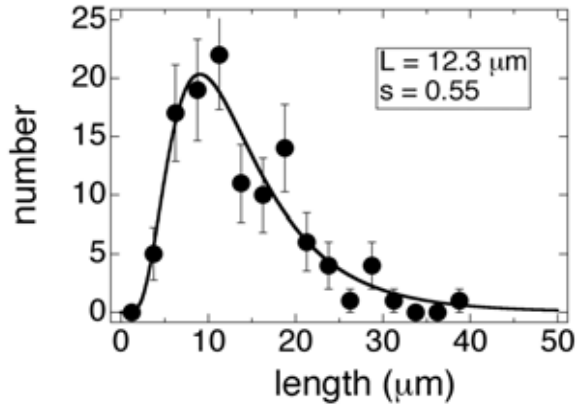

*Figure 5 : Probability distribution function of length for the rods seen in Figs. 4. The continuous line was derived from best fit calculation using a log-normal distribution. For this dispersion, the average length of the rods was found to be 12.3 μm, and the polydispersity s = 0.55.*

Our objective is now to demonstrate that these observations are consistent with the behavior of superparamagnetic rods. Superparamagnetic here implies that the rods do not carry a permanent magnetic moment (in contrast to ferromagnetic materials [29]), but acquire one under the application of a field. The reorientation dynamics of various kinds of superparamagnetic rods in conditions similar to those of the experiment were described in details earlier [12, 21, 28]. When inertial terms are neglected [12], the rod relaxation results from the balance between the magnetic and the hydrodynamic torques, yielding a differential equation for the time evolution $\theta(t)$ :

$$\frac{d\theta}{dt} = -\frac{k}{2}\sin(2\theta) \quad (1)$$

$$\text{where } k = \frac{\chi^2}{(2+\chi)}\frac{\mu_0\, g(L_{Rod}/D_{Rod})}{2\eta} D_{Rod}^2 \left(\frac{H}{L_{Rod}}\right)^2 \quad (2)$$

In Eq. 2, $\mu_0$ denotes the permeability of vacuum, $\chi$ the magnetic susceptibility of the material, $\eta$ the viscosity of the solvent, $D_{Rod}$ and $L_{Rod}$ diameter and length of the rod. The function $g(L_{Rod}/D_{Rod})$ is a slowly varying function of the aspect ratio $L_{Rod}/D_{Rod}$ [21]. The solution of the differential equation in the interval $[0, \pi]$ reads :

$$\ln|tg\,\theta(t)| = -kt + C \quad (3)$$

where C is a constant of integration. With the initial conditions considered here ($\theta_I = \pi/2$ at $t = 0$), the integration constant C becomes infinite, and the solution of the differential equation expresses simply as :

$$\theta(t) = \frac{\pi}{2} = cste \quad (4)$$

In other terms, according to the above model, if the initial angle is exactly equal to $\pi/2$, the rods do not move and remain in their original position. This is equivalent to say that in this orientation, the magnetic torque is zero. $\theta_I = \pi/2$ represents an equilibrium orientation that is mechanically unstable, and this is in part the reason why the rods seem to "hesitate" at the application of the field.

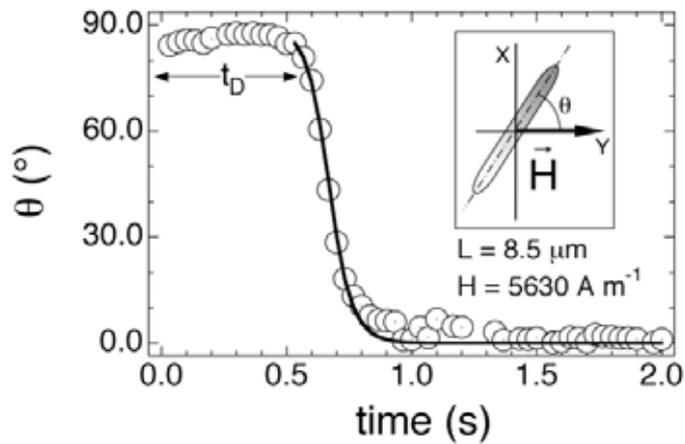

***Figure 6*** *: Time dependence of the angle θ(t) between the rod and applied magnetic excitation H. Note that during the delay time $t_D$, the rod remains close to the initial orientation at $\theta_I = \pi/2$. In the experiment shown, the rod had a length of $L_{Rod} = 8.5$ μm and the applied field was 5630 A m$^{-1}$. The continuous line was obtained through Eq. 5 (see text). Inset : Schematic representation of the reorientation of the rod initially along the X-axis and subjected to a field applied along the Y-axis.*

Thanks to the Brownian motions, as time is running, the rods explored orientation angles that are away from the initial $\pi/2$-orientation. If the excursions are large enough (a few degrees are usually necessary), the net magnetic torque can become sufficient to rotate the rod up to its final orientation. Obviously, the model based on the balance between the magnetic and hydrodynamic torques is oversimplified, and does not take into account the rotational diffusion of the rods due to thermal motions. The present results suggest that the Brownian reorientations around $\theta_I$ help to trigger the onset of rotation. These results also allow to understand why the delay time $t_D$ can vary from an experiment to another, from few milliseconds to seconds. Above this delay time however, the time dependence for tg$\theta(t)$ can be predicted accurately. It reads [21] :

$$\text{tg}\,\theta(t) = \text{tg}\,\theta_0 \exp\left[-k(t - t_D)\right] \quad (4)$$

Here $\theta_0$ is some initial angle, defined by convenience at $t_D$, so that $\theta_0 = \theta(t_D)$. In Eq. 5, for $\theta$ and $\theta_0$ belonging to the first quadrant, the tangents are positive, and as such the angle $\theta(t)$ decreases from $\theta = \pi/2$ to 0. The rod rotates clockwise. For the angles belonging to the second quadrant, the tangents are negative, and there the angle $\theta(t)$ increases from $\pi/2$ to $\pi$ in a counterclockwise motion. The data in Fig. 6 and in Figs. 7 were fitted using Eq. 5 for angles in the first quadrant, and with the values for k and $\theta_0$ listed in Table I. The good agreement between the experimental data and the predictions of Eq. 5 (continuous lines in the two figures) attests of the reliability of the model. In Ref. [21], the quadratic dependence of the parameter k (Eq. 2) was demonstrated by plotting this parameter as a function of the ratio $H/L_{Rod}$.

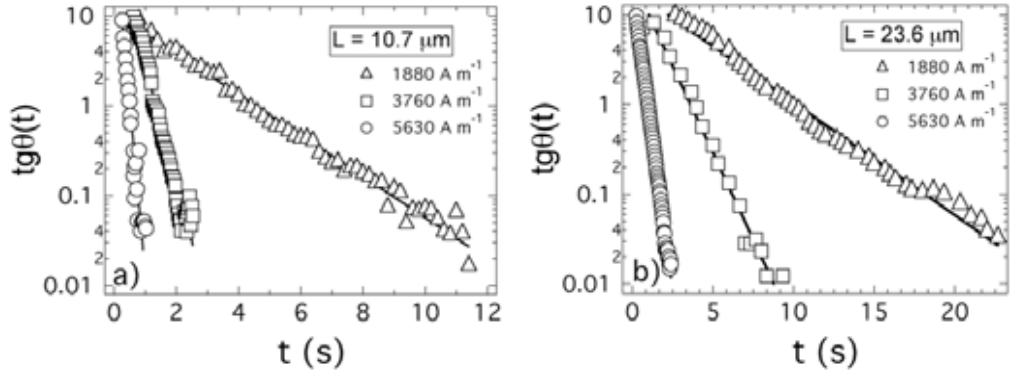

*Figure 7 : Tangent of the reorientation angle θ(t) as a function of the time for two rods of length 10.7 μm (a) and 23.6 μm (b). Note that the exponential decay was observed over 3 decades in the ordinate units, and that the decay rate k (Eq. 2) strongly depends on the applied magnetic excitation. The straight lines are from Eq. 5, with parameters listed in Table I*

|  | H (A m$^{-1}$) | $t_D$ (s) | $\theta_0$ (°) | k (s$^{-1}$) |
|---|---|---|---|---|
| rod in Fig. 7a $L_{Rod} = 10.7$ μm | 1880 | 1.0 | 84 | 0.6 |
|  | 3760 | 0.6 | 86 | 3 |
|  | 5630 | 0.1 | 88 | 8 |
| rod in Fig. 7b $L_{Rod} = 23.6$ μm | 1880 | 1.5 | 85 | 0.25 |
|  | 3760 | 0.3 | 86 | 0.9 |
|  | 5630 | 0.2 | 87 | 3 |

***Table I*** : *List of parameters determined from the data in Figs. 7 and using Eq. 5.*

# IV – Conclusion

In a previous paper [21], we had shown that the electrostatic interactions between oppositely charged magnetic nanoparticles and polymers can be accurately controlled, and that under appropriate conditions, this control resulted in the fabrication of nanostructured rods in the micrometer range. Remarkable properties unveiled in this first report were large aspect ratios and significant stiffness. In the present paper, we focused on the rod behaviors under the application of a magnetic field. Reorientation kinetics for fields applied perpendicular to the initial rod orientation have shown three main properties. The rods "hesitate" before undergoing the rotation during a delay time that span between few milliseconds and few seconds. The rotation occurs with an equi-probability in the clockwise and counterclockwise directions. Finally, the long time decrease is such that the tangent of the rotation angle decreases exponentially with time. We have provided here a simple model that account for these three properties, allowing us to conclude that the rods have inherited the properties of the single maghemite nanoparticles, namely to be superparamagnetic. The present approach of nanoparticle co-assembly should open new perspectives for the fabrication of nanodevices such as tips, tweezers and actuators applicable in biophysics and biomedecine.

**Acknowledgements** : The authors are indebted to Annie Vacher and Marc Airiau (Centre de Recherches d'Aubervilliers, Rhodia, France) for the cryo-TEM experiments on the nanoparticles. We are grateful to Jean-Paul Chapel from the Complex Fluids Laboratory (CRTB Rhodia, Bristol, Pa, USA) for fruitful discussions and to Benoit Ladoux from the Laboratoire Matière et Systèmes Complexes (Université Paris-Denis Diderot) for access to microscopy and imaging facility. Parts of the research performed here was financed by the Agence Nationale de la Recherche, under the contract BLAN07-3_206866 and by Rhodia.